\documentclass[useAMS,usenatbib,usegraphicx]{mn2e}
\def\LB1{L1157--B1}

\usepackage{fixltx2e}
\title[HDO emission around SVS13-A]{Hot and dense water in the inner 25 AU of SVS13-A}
\author[C. Codella et al.]{C. Codella$^{1}$\thanks{E-mail:
codella@arcetri.astro.it },  
C. Ceccarelli$^{2,3,1}$, E. Bianchi$^{1,4}$, L. Podio$^{1}$, R. Bachiller$^{5}$, 
B. Lefloch$^{2,3}$, \newauthor F. Fontani$^{1}$, V. Taquet$^{6}$, L. Testi$^{1,7}$ 
\\
\\
$^{1}$ INAF-Osservatorio Astrofisico di Arcetri, L.go E. Fermi 5, Firenze, 50125, Italy \\
$^{2}$ Univ. Grenoble Alpes, IPAG, F-38000 Grenoble, France \\
$^{3}$ CNRS, IPAG, F-38000 Grenoble, France \\
$^{4}$ Dipartimento di Fisica e Astronomia, Universit\`a degli Studi di Firenze, Italy \\
$^{5}$ IGN, Observatorio Astron\'omico Nacional, Calle Alfonso XII, 28004 Madrid, Spain \\ 
$^{6}$ Leiden Observatory, Leiden University, 9513, 2300-RA Leiden, The Netherlands \\
$^{7}$ ESO, Karl Schwarzchild Srt. 2, 85478 Garching bei M\"unchen, Germany
}

\begin{document}

\date{Accepted date. Received date; in original form date}

\pagerange{\pageref{firstpage}--\pageref{lastpage}} \pubyear{2011}

\maketitle

\label{firstpage}

\begin{abstract}
In the context of the ASAI (Astrochemical Surveys At IRAM)
project, we carried out an unbiased spectral survey in the 
millimeter window towards the well known 
low-mass Class I source SVS13-A.
The high sensitivity reached (3--12 mK) allowed us 
to detect at least 6 
HDO broad (FWHM $\sim$ 4--5 km s$^{-1}$) emission lines 
with upper level energies up to $E_{\rm u}$ = 837 K.
A non-LTE LVG analysis implies the presence 
of very hot 
(150--260 K) and dense ($\geq$ 3 $\times$ 10$^7$ cm$^{-3}$) gas 
inside a small radius ($\sim$ 25 AU) around the star, 
supporting, for the first time, the occurrence of
a hot corino around a Class I protostar. 

The temperature is higher than expected for 
water molecules are sublimated
from the icy dust mantles ($\sim$ 100 K). 
Although we cannot exclude we are observig the effects of 
shocks and/or winds at such small scales,
this could imply that the observed HDO
emission is tracing 
the water abundance jump expected at temperatures
$\sim$ 220--250 K, when the activation barrier of the 
gas phase reactions leading to the formation of water can be overcome. 
We derive $X(HDO)$ $\sim$ 3 $\times$ 10$^{-6}$, and a H$_2$O deuteration
$\geq$ 1.5 $\times$ 10$^{-2}$, suggesting that
water deuteration does not decrease as the protostar  
evolves from the Class 0 to 
the Class I stage.
\end{abstract}


\section{Introduction}\label{sec:introduction}

The origin of terrestrial water is still a source of intense debate
(e.g. Ceccarelli et al. 2014; van Dischoeck et al. 2014; Altwegg et
al. 2015).  A key element to shed light on it, is how water evolves
with time in proto-Sun analogues.  Specifically, two aspects are
particulary important: (1) the amount of water and its distribution in
the planet formation region (a few tens of AU) of the proto-Suns, and
(2) its deuterium fractionation (e.g. Ceccarelli et al. 2014; Willacy
et al. 2015).

With respect to the first point, water has been detected at all stages
of the Sun-like star formation process, from prestellar cores and
Class 0 sources to the Solar System (e.g. Caselli et al. 2012;
Ceccarelli et al. 1999; van Dishoeck et al. 2011, 2014). However, so
far, we mostly have poor angular resolution observations that allowed
us to detect the water emission, but not to resolve it on small
($\le 1000$ AU) scales. Only a handful of observations exists with
enough spatial resolution. They show that the water emission in the
envelopes of Class 0 sources is concentrated in small regions, called
hot corinos (Codella et al. 2010; Persson et al. 2013, 2014; Taquet et
al. 2013; Coutens et al. 2014). On the contrary, no spatially resolved
observations exist for the more evolved protostars, the Class I
sources. With respect to the water deuterium fractionation, again only
a few measures are available in Class 0 sources (Coutens et al. 2012,
2013, 2014; Persson et al. 2013, 2014; Taquet et al. 2013), but none
in Class I.

In the context of the ASAI (Astrochemical Surveys At IRAM: {\it
  http://www.oan.es/asai/}) project, we have carried out a systematic
study of the molecular emission towards SVS13-A. This is a well
studied young stellar object located in the NGC1333 star-forming
region, at a distance of $\sim$235 pc (Bachiller et al. 1998; Hirota
et al. 2008; Lee et al. 2016). SVS13-A is part of the system SVS13,
where three millimeter sources have been identified by interferometric
observations (Bachiller et al. 1998; Looney et al. 2000), called A, B,
and C. The distance between A and B is 15$\arcsec$, while C is 
20$\arcsec$ away from A. The systemic velocity  
of the sources A and B is between +8 km s$^{-1}$ and 
+9 km s$^{-1}$ (Chen et al. 2009;
Lop\`ez-Sepulcre et al. 2015). The luminosity of SVS13-A has been
estimated to be 34 L$_\odot$ (Chen et al. 2009; Tobin et al. 2016), 
where we corrected
for the new estimate of the distance $d$ = 235 pc (Hirota et al. 2008).

Although SVS13-A is still deeply embedded in a large-scale ($\sim$6000
AU; e.g. Lefloch et al. 1998) envelope, its extended ($>$0.07 outflow
pc) outflow, associated with the HH7-11 chain (e.g. Lefloch et
al. 1998, and references therein), and its low L$_{submm}$/L$_{bol}$
ratio ($\sim$ 0.8 $\%$) lead to the classification as a Class I source
(e.g. Chen et al. 2009 and references therein). 

In this Letter, we report the detection of several lines of HDO
towards SVS13-A, providing the first detection of deuterated water in
a Class I source.

\section{Observations}\label{sec:observations}

The present observations have been performed during several runs
between 2012 and 2014 with the IRAM 30-m telescope near Pico Veleta
(Spain) in the context of the Astrochemical Surveys At
IRAM\footnote{www.oan.es/asai} (ASAI) Large Program.  In particular,
the unbiased spectral surveys at 3 mm (80–116 GHz), 2 mm (129–173
GHz), and 1.3 mm (200–276 GHz) have been acquired using the EMIR
receivers with a spectral resolution of 0.2 MHz.  The observations
were performed in wobbler switching mode with a throw of 180$\arcsec$
towards the coordinates of the SVS13-A object, namely
$\alpha_{\rm J2000}$ = 03$^{\rm h}$ 29$^{\rm m}$ 03$\fs$29,
$\delta_{\rm J2000}$ = +31$\degr$ 16$\arcmin$ 03$\farcs$8).  The
pointing was checked by observing nearby planets or continuum sources
and was found to be accurate to within 2$\arcsec$--3$\arcsec$.  The
HPBWs are in the 9$\arcsec$--31$\arcsec$ range.

The data were reduced with the the
GILDAS--CLASS\footnote{http://www.iram.fr/IRAMFR/GILDAS} package.
Calibration uncertainties are $\simeq$ 10\% at 3mm and $\sim$ 20\% at
shorter wavelengths. All the spectra have been converted from antenna
temperature to main beam temperature ($T_{\rm MB}$), using the main
beam efficiencies reported on the IRAM 30-m
website\footnote{http://www.iram.es/IRAMES/mainWiki/Iram30mEfficiencies}.

\section{Results and discussion}

\subsection{HDO detected lines}\label{sec:hdo-detected-lines}

The ASAI unbiased survey allows us to detect 7 HDO lines (1 in the 3
mm band, 1 in the 2 mm band, and 5 in the 1.3 mm one) covering a wide
range of excitation, with upper level energies E$_{\rm u}$ from 47 K
to 837 K. 
The 81 GHz line is only tentatively detected, given
the low S/N ratio. However, the following analysis will show
how the 1$_{1,0}$--1$_{1,1}$ intensity is well in agreement
with those of the other lines observed at 2mm and 1.3mm.
The 3$_{3,1}$--4$_{2,2}$ transition (at $\sim$ 242.0 GHz) is
the HDO line with the highest upper level energy (E$_{\rm u}$=837 K)
ever observed towards a low-mass protostar.  The profiles of all
detected lines are shown in Fig. 1, while Table 1 reports the results
of the line Gaussian fits.  The HDO emission peaks at the cloud
systemic velocity, between +8.0 and +9.0 km s$^{-1}$ (Chen et al. 2009;
Lop\`ez-Sepulcre et al. 2015). The lines are quite broad, with a
$FWHM$ $\sim$ 4.2--4.9 km s$^{-1}$ for all the lines but the two
observed at 3mm and 2mm, which are also those observed with the lowest
S/N ratio and the worst spectral resolution (from 0.7 km s$^{-1}$ to
0.2 km s$^{-1}$ moving from 80.6 GHz and 266.2 GHz).
The emission in the 1 and 2 mm bands only originates from SVS13-A, as
SVS13-B, 15$\arcsec$ south-west, is outside the $HPBW$. The 3 mm band
might, in principle, contain some emission from SVS13-B. However, the
analysis of the measured fluxes tends to exclude a substantial
contamination from SVS13-B also in this band (see below).

Finally, we searched for H$_2^{18}$O lines in our spectral survey and
found none. The most sensitive upper limit to the H$_2$O column
density is set by the non detection of the para--H$_2^{18}$O
3$_{\rm 1,3}$--2$_{\rm 2,0}$ line at 
203.40752 GHz\footnote{from  
Jet Propulsion Laboratory database, 
http://spec.jpl.nasa.gov/home.html; Pickett et al. (1998)}. We
obtained a 3$\sigma$ upper limit on the peak temperature (in
$T_{\rm MB}$ scale) of 20 mK.

\begin{figure}
\centerline{\includegraphics[angle=0,width=8cm]{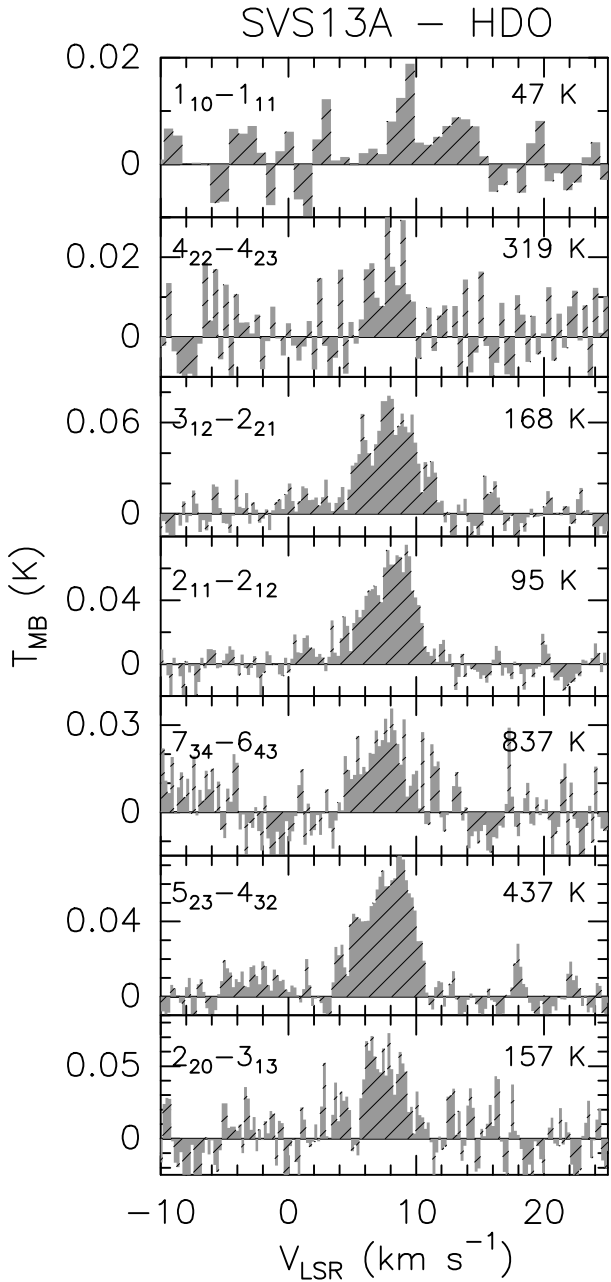}}
\caption{HDO line profiles (in main-beam temperature, $T_{\rm MB}$,
  scale) observed with the IRAM 30-m antenna towards SVS13-A. In each
  panel, both the transition and the upper level energy, $E_{\rm u}$,
  are reported.}
\label{spectra}
\end{figure} 

\begin{table*}
% \centering
% \begin{minipage}{140mm}
  \caption{List of HDO transitions and line properties (in $T_{\rm MB}$ scale) detected towards SVS13-A}
  \begin{tabular}{lccccccccccc}
  \hline
\multicolumn{1}{c}{Transition} &
\multicolumn{1}{c}{$\nu$$^{\rm a}$} &
\multicolumn{1}{c}{$HPBW$} &
\multicolumn{1}{c}{$g_{\rm u}$$^a$} &
\multicolumn{1}{c}{$E_{\rm u}$$^a$} &
\multicolumn{1}{c}{$S\mu^2$$^a$} &
\multicolumn{1}{c}{log(A/s$^{-1}$)$^a$} &
\multicolumn{1}{c}{rms} &
\multicolumn{1}{c}{$T_{\rm peak}$$^b$} &
\multicolumn{1}{c}{$V_{\rm peak}$$^b$} &
\multicolumn{1}{c}{$FWHM$$^b$} &
\multicolumn{1}{c}{$I_{\rm int}$$^b$} \\
\multicolumn{1}{c}{ } &
\multicolumn{1}{c}{(GHz)} &
\multicolumn{1}{c}{($\arcsec$)} &
\multicolumn{1}{c}{} &
\multicolumn{1}{c}{(K)} &
\multicolumn{1}{c}{(D$^2$)} & 
\multicolumn{1}{c}{} &
\multicolumn{1}{c}{(mK)} &
\multicolumn{1}{c}{(mK)} &
\multicolumn{1}{c}{(km s$^{-1}$)} &
\multicolumn{1}{c}{(km s$^{-1}$)} &
\multicolumn{1}{c}{(mK km s$^{-1}$)} \\ 
\hline
1$_{\rm 1,0}$--1$_{\rm 1,1}$ &  80.57829 & 31 & 3 & 47 & 0.66 &  --5.88 &  5 & 14(2) & +9.2(0.3) & 1.7(0.9) &  26(9) \\
4$_{\rm 2,2}$--4$_{\rm 2,3}$ & 143.72721 & 17 & 9 & 319 & 0.73 & --5.55 &  9 & 21(8) & +8.0(0.3) & 3.1(0.8) & 68(20) \\
3$_{\rm 1,2}$--2$_{\rm 2,1}$ & 225.89672 & 11 & 7 & 168 & 0.69 & --4.88 &  7 & 45(7) & +8.0(0.1) & 4.9(0.3) & 234(16) \\
2$_{\rm 1,1}$--2$_{\rm 1,2}$ & 241.56155 & 10 & 5 & 95 & 0.36 &  --4.92 &  6 & 41(6) & +8.0(0.1) & 4.7(0.3) & 206(9) \\
7$_{\rm 3,4}$--6$_{\rm 4,3}$ & 241.97357 & 10 & 15 & 837 & 1.39 & --4.82 &  7 & 17(5) & +7.7(0.3) & 4.6(0.7) & 83(10) \\
5$_{\rm 2,3}$--4$_{\rm 3,2}$ & 255.05026 & 10 & 11 & 437 & 1.02 & --4.75 &  6 & 40(6) & +7.8(0.1) & 4.5(0.2) & 199(9) \\
2$_{\rm 2,0}$--3$_{\rm 1,3}$ & 266.16107 & 9  & 5 & 157 & 0.40 & --4.76 & 12 & 31(9) & +7.4(0.3) & 4.2(0.8) & 141(19) \\
\hline
\end{tabular}
%\end{minipage}

$^a$ From the Jet
Propulsion Laboratory database (Pickett et al. 1998). 
$^b$ The errors are the gaussian fit uncertainties. \\
\end{table*}

\subsection{Analysis of the HDO emission}\label{sec:analys-hdo-emiss}

We analysed the observed HDO line emission with the non-LTE LVG model
by Ceccarelli et al. (2003), using the collisional coefficients for
the system HDO-H$_2$ computed by Faure et al. (2012) and extracted
from the the BASECOL database (Dubernet et al. 2013). We assumed a
plane-parallel geometry and a Boltzmann distribution for the
ortho-to-para H$_2$ ratio of 3. 
Note that the collisional
coefficients with ortho-H$_2$ can be a factor 5 larger than the
corresponding coefficients with para-H$_2$ (Faure et al. 2012), but
only at low temperatures ($\ll$45 K) and not at those here discussed
(see below).  Note also that the HDO 7$_{\rm 3,4}$--6$_{\rm 4,3}$ line (with
$E_{\rm u}$ = 837 K) has been excluded from the LVG analysis because
the corresponding collisional rates have not been calculated (see
later for a comparison with an LTE approach).

\begin{figure}
%\centerline
\begin{center}
\includegraphics[angle=0,width=7.6cm]{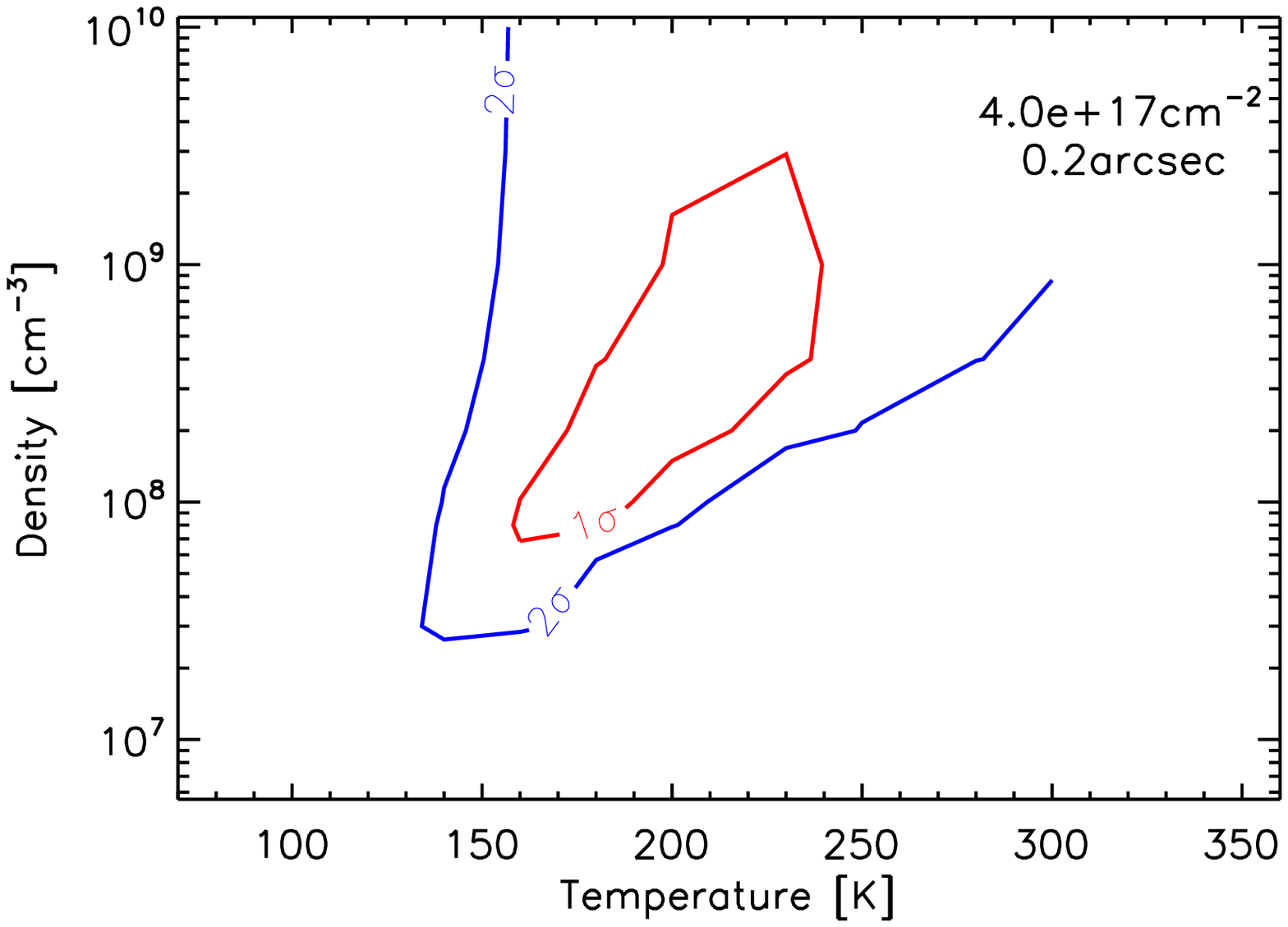}\\
\includegraphics[angle=0,width=7.7cm]{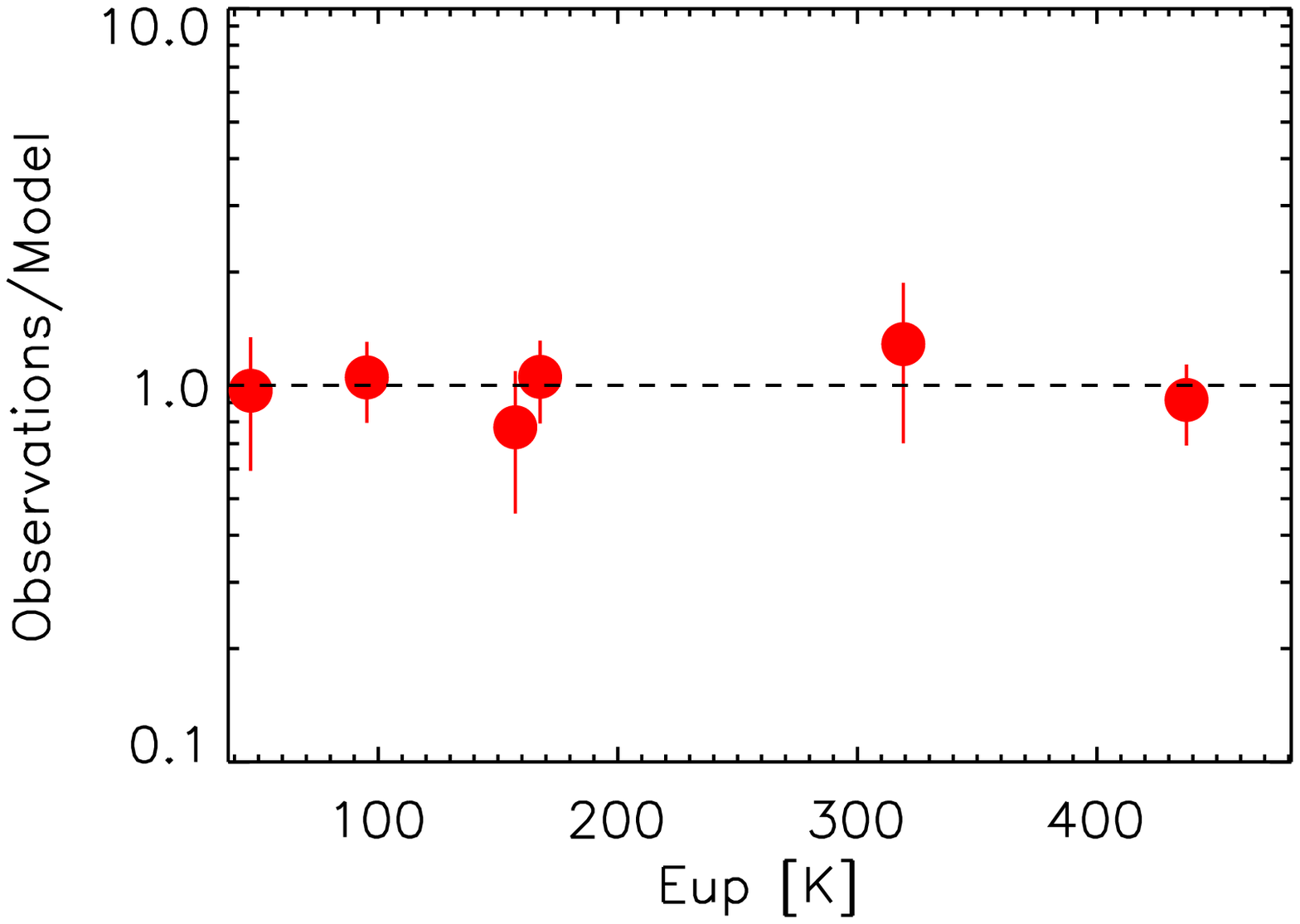}
\end{center}
\caption{{\it Upper panel:} The $\chi^2_{\rm
    r}$ contour plot obtained considering the non-LTE model predicted
  and observed intensities of all detcted HDO lines but the 241.974
  GHz line with $E_{\rm u}$ = 837 K (for which no collisional rates
  are available).  The best fit is obtained with 0$\farcs$2, $N_{\rm
  HDO} = 4 \times 10^{17}$ cm$^{-2}$, $T_{\rm kin} = 200$ K,
   and $n_{\rm H_2} = 2 \times 10^{8}$ cm$^{-3}$).
  The 1$\sigma$  and 2$\sigma$  of the $\chi^2_{\rm r}$ contours are
  reported.  {\it Lower panel:} Ratio between the observed line
  intensities with those predicted by the best fit model as a
  function of line upper level energy $E_{\rm u}$.}
\label{lvg}
\end{figure}

We run a large grid of models varying the temperature $T_{\rm kin}$
from 100 to 300 K, the H$_2$ density $n_{H2}$ from $8\times 10^8$ to
$1\times 10^{10}$ cm$^{-3}$, the HDO column density $N(HDO)$ from
$4\times 10^{16}$ to $7\times 10^{17}$ cm$^{-3}$, and the emitting
sizes $\theta_s$ from 0.05 to 10 $\arcsec$. The lowest
$\chi^2_{\rm r}$ is obtained with $N(HDO)=4\times10^{17}$ cm$^{-2}$,
and $\theta_s=$0$\farcs$2, corresponding to $\sim50$ AU.
Figure \ref{lvg} (upper panel) shows the $\chi^2_{\rm r}$ contour plot
as a function of the temperature and H$_2$ density with these values.
The best fit solution is found for a very high temperature,
$T_{\rm kin}$=150--260 K, and a quite high density
$n_{\rm H_2} \geq 3 \times 10^7$ cm$^{-3}$.  Figure \ref{lvg} (lower
panel) shows the goodness of the fit, namely the ratio between the
measured velocity-integrated intensities and the LVG model
predictions, as a function of the line upper level energy, for the
best fit solution: $N(HDO)=4\times10^{17}$ cm$^{-2}$,
$\theta_s=$0$\farcs$2, $T_{\rm kin} = 200$ K, and
$n_{\rm H_2} = 2 \times 10^8$ cm$^{-3}$. The lines are predicted to
be optically thin to moderately thick. The largest opacities are
$\sim2$ for the four lowest lying lines (at 80.58, 241.56, 266.16 and
225.90 GHz), while the other lines have opacities lower than
unity. 

Finally, the populations of the detected transitions are
predicted to be close to LTE. 
Indeed, since non-LTE predictions were not
possible for the higher lying line at 241.97 GHz, we also computed 
the LTE solution (see Fig. 3), finding a 
rotational temperature of 334$\pm$42 K, which is larger than
the kinetical temperature derived from the LVG analysis, i.e. 160--240 K
considering the 1$\sigma$ $\chi^2_{\rm r}$ solution.
In practice, assuming the non-LTE LVG solution, the agreement  
between the predicted and observed intensity 
of the $E_{\rm u}$ = 857 K is within a
factor 2; this is acceptable 
if we consider the opacity of the
low-$E_{\rm u}$ lines and that the LTE condition may not apply to such 
a high lying line so that we cannot exclude a temperature gradient 
with a component with $T_{\rm kin}$ larger than 200 K. High spatial
resolution observations are required to clarify this point.

\subsection{Origin of the HDO emission}\label{sec:origin-hdo-emission}

The non-LTE analysis reveals the presence of a hot (150--260 K), dense
($\geq 3\times 10^7$ cm$^{-3}$) and compact ($\sim$ 50 AU) component in
SVS13-A. 
Note that the density estimate well agrees with
that derived from dust continuum by Chen et al. (2009). 
High temperatures (218 K) have been similarly obtained by Coutens et al. (2014)
applying an LTE analysis to four HDO lines detected, using the PdB array,
towards the Class 0 NGC1333 IRAS2A. 
Also, high excitation conditions have been found through LVG analysis of H$_2$O
lines as observed by Herschel towards both Class 0 and
Class I sources 
(e.g. Herczeg et al. 2012; Podio et al. 2012; Karska et al. 2013; Busquet et al. 2014;
Kristensen et al. 2016);
in these cases, the high excitation H$_2$O emission has been associated with shocked gas induced by jets.
Thus, the present results may indicate the presence of 
jet-induced shocks on the $\sim$ 20 AU scales. 
However, the jet hypothesis is ruled out by:                     
(i) the $FWHM$ line profiles ($\simeq$ 4 km s$^{-1}$)
(the is usually traced by velocities larger than 10 km s$^{-1}$),
and (ii) the compact size inferred by the LVG analysis.
On the other hand, HDO emission could be emitted by
shocks induced by the viscosity of a disk or it
could probe the base of a disk wind (as
suggested by Codella et al. 2016 for the Class 0 HH212).
Again, there is no signature in the line profile suggesting
an association of water with a small
portion of the large disks expected to be around Class I objects
(around 250 AU; e.g. Eisner et al. 2012).

In conclusion, what is HDO tracing in the SVS13-A system? 
The observed high temperature and linewidth are
consistent with {\it the presence of a hot corino} inside SVS13-A,
where the gas is thermally heated by the central source. Of
course, the definition of hot corino involves the detection of complex
organic molecules (e.g. Ceccarelli et al. 2007). We anticipate that
this is indeed the case for SVS13-A (Bianchi et al., in preparation).
Assuming that the dust is heated
by the 34 L$_\odot$ central source and that the dust emission is
optically thin, the dust temperature would be about 200 K at a
distance of $\sim$ 25 AU (see e.g. Ceccarelli et al. 2000, Eq. 1). Of
course, if the dust opacity is thick in the innermost regions, then
this value is a lower limit. Therefore, this temperature is in good agreement
with the LVG analysis. 
Indeed, high temperatures from HDO were observed by 
Coutens et al. (2014) towards the Class 0 object NGC1333-IRAS2A, 
in agreement with the present hypothesis that HDO lines, 
being optically thin, probe inner regions around
the protostars.

However, the hot corino interpretation has a problem. If the high
temperature is caused by the thermal heating, one would expect that
water is sublimated from the icy grain mantles at $\sim$ 100 K, whereas
the HDO line emission indicates a larger temperature. Why?
A first possibility could be that the HDO line emission is dominated
by warmer gas because of the line opacities. Indeed, even if the HDO
abundance has a jump at $\sim$ 100 K, if the lines are optically thin
then the warmer regions, with higher opacities may dominate the
integrated intensity. One has also to consider that the water
abundance has a further jump at around 220--250 K, caused by the
reactions that convert all the gaseous atomic oxygen into water and
that possess activation barriers making them efficient at $\geq$ 220 K
(Ceccarelli et al. 1996). The temperature is close to that derived
from the LVG modeling, so that it is a plausible hypothesis that the
gas probed by the observed HDO lines lies in a region warmer than the
water desorption region because of the line opacities.

\begin{figure}
\centerline{\includegraphics[angle=90,width=8cm]{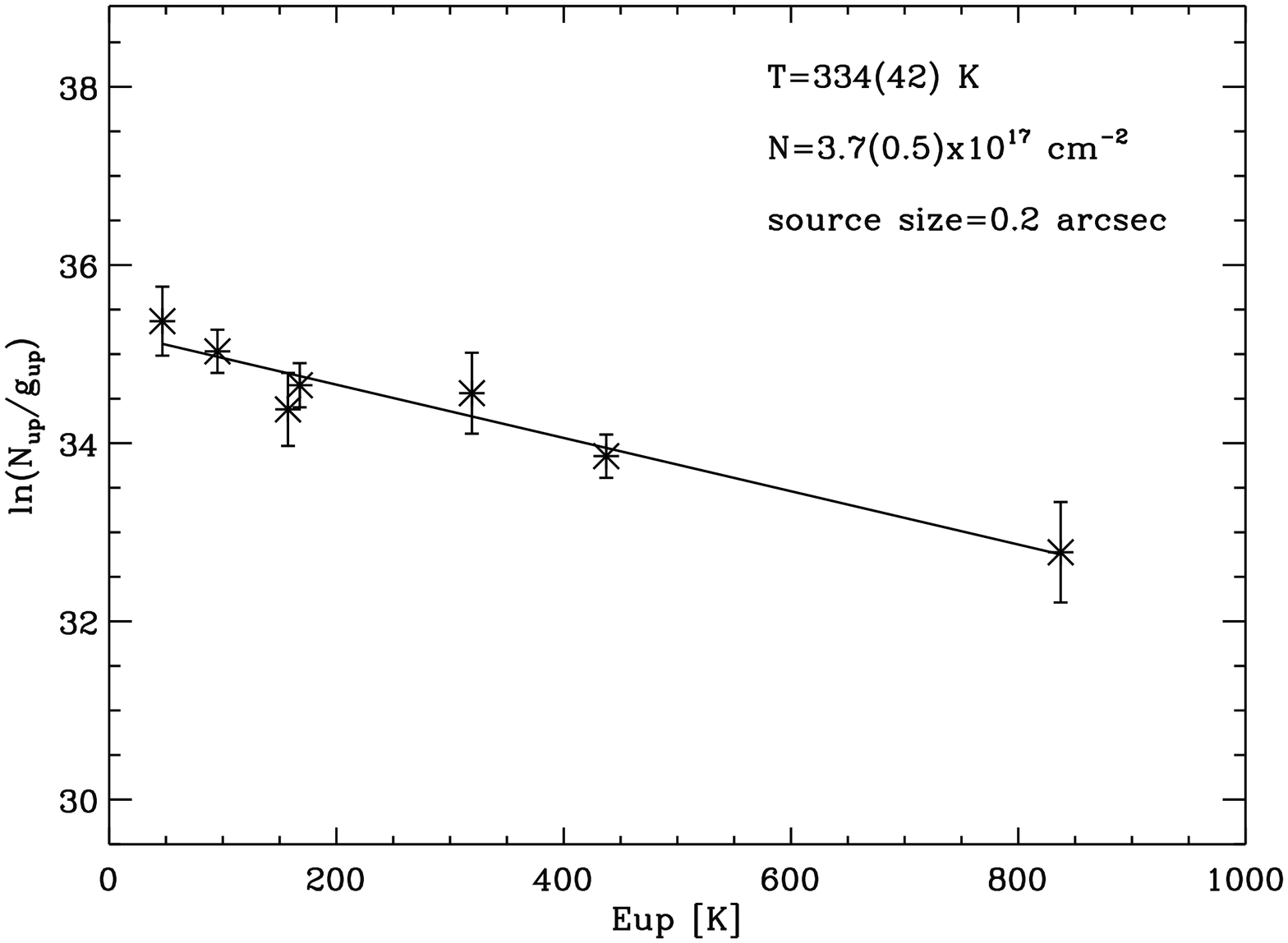}}
\caption{Rotation diagram for the
HDO transitions reported in Table 1, including the $E_{\rm up}$ = 837 K,
not considered in the LVG analysis (see Fig. 2). The parameters
$N_{\rm up}$, $g_{\rm up}$, and $E_{\rm up}$ are respectively
the column density, the degeneracy, and the energy of the upper level.
The error bars on ln($N_{\rm up}$/$g_{\rm u}$) are
given by the vertical bar of the symbols.
The plot allows us to derive a rotational temperature of
334$\pm$42 K and a total column density of
3.7$\pm$0.5 $\times$ 10$^{17}$ cm$^{-2}$.}
\label{rd}
\end{figure}

\subsection{Water deuteration}\label{sec:water-deuteration}

Using the intensity 3$\sigma$ upper limit of the para--H$_2^{18}$O
line at 203.40752 GHz (see \S3.1) and
assuming a source size of 0$\farcs$2 and a temperature of 200 K, 
we derive un upper limit to the H$_2^{18}$O column density of
$N(H_2^{18}O) \leq 8 \times 10^{17}$ cm$^{-2}$.  Assuming the standard
value of $^{16}$O/$^{18}$O = 560, the upper limit to the water column
density is $N(H_2O) \simeq 4 \times 10^{20}$ cm$^{-2}$.
Using the HDO column density previously derived,
$N(HDO)$ = 4 $\times$ 10$^{17}$ cm$^{-2}$, leads to a lower limit to the
water deuteration, $\geq$ 1 $\times$ 10$^{-3}$. 
Finally, using the derived $n_{H_2}$ density (2 $\times$ 10$^8$ cm$^{-3}$)
and emitting sizes (0$\arcsec$2 = 50 AU in diameter) provides an
estimate of the H$_2$ column density of $\sim$ 1.5 $\times$ 10$^{23}$
cm$^{-2}$ and, consequently of the HDO abundance of $\sim$ 3 $\times$
10$^{-6}$. Similarly, the upper limit to the H$_2$O column density can
be converted into an upper limit to the water abundance, namely $\leq$
3 $\times$ 10$^{-3}$. We can, therefore, increase the real lower limit to the
water deuteration considering that, reasonably, the water abundance cannot be
larger than about $\sim$ 2 $\times$ 10$^{-4}$. This leads to a lower limit
HDO/H$_2$O $\geq$ 0.015.

This upper limit is consistent with those derived so far towards Class
0 protostars, $\sim$ 10$^{-2}$ by Coutens et al. (2012) 
and Taquet et al. (2013), 0.3--8 $\times$ 10$^{-2}$, and 
larger than those quoted by Persson et
al. (2013, 2014) and Coutens (2013; 2014), 0.1--4 $\times$ 10$^{-3}$. 
Therefore, the deuteration of water does not seem to
diminish from Class 0 to Class I sources. Yet, we conclude with a word
of prudence, as this value of deuteration has been obtained taking the values
of the LVG modeling. In particular, since the lines seem close to the
LTE, the H$_2$ density could be larger and, consequently, the HDO
abundance could be lower by the same factor.

\section{Conclusions}

The high-sensitivity of the IRAM 30-m ASAI unbiased spectral survey 
in the mm-window allows 
us to detect towards the Class I object SVS13-A a large number
of HDO emission lines with upper level energies up to $E_{\rm u}$ = 837 K.
The non-LTE LVG analysis points to hot 
(150--260 K), dense ($\geq$ 3 $\times$ 10$^7$ cm$^{-3}$) gas associated with
a quite small emitting region (50 AU), supporting the occurrence of
a hot corino inside SVS13-A. The HDO abundance is found to be
$\sim$ 3 $\times$ 10$^{-6}$.
Although the occurrence of shocks at such small
scales cannot be excluded, it is tempting to suggest we are observing 
for the first time the jump in the water abundance occurring at temperatures 
higher than 200 K, when the activation barriers of the gas phase reactions
coverting oxygen into water can be overcome.   

Obviously,
the final answer is in the hands of future interferometric  
observations (e.g. using ALMA) imaging water emission around SVS13-A
on scales $\leq$ 20 AU.

\section*{Acknowledgments}

The authors are grateful 
to the IRAM staff for its help in the
calibration of the PdBI data. 
We also thank F. Dulieu for instructive discussions. 
The research leading to these results
has received funding from the European Commission Seventh Framework
Programme (FP/2007-2013) under grant agreement N° 283393 (RadioNet3).
This work was partly supported by the PRIN INAF 2012 -- JEDI and by
the Italian Ministero dell'Istruzione, Universit\`a e Ricerca through
the grant Progetti Premiali 2012 -- iALMA. CCe \& BL acknowledge the
financial support from the French Space Agency CNES, and RB
from Spanish MINECO (through project FIS2012-32096).

{}

\bsp

\label{lastpage}

\end{document}